\documentclass[prc,twocolumn,superscriptaddress]{revtex4-1}
\usepackage{gensymb}
\usepackage{amsmath,amssymb,epsfig,bm,mathrsfs,color}
\usepackage{slashed}
\usepackage{color}
\newcommand{\be}{\begin{equation}}
\newcommand{\ee}{\end{equation}}
\newcommand{\bea}{\begin{eqnarray}}
\newcommand{\eea}{\end{eqnarray}}

\newcommand{\aap}{Astron.\& Astrophys.}

\newcommand{\vecnabla}{{\bm \nabla}}

\newcommand{\vecA}{{\bm A}}
\newcommand{\vecB}{{\bm B}}
\newcommand{\vecj}{{\bm j}}

\newcommand{\mnras}{{Mon. Not. R. Astron. Soc.}}
\newcommand{\physrep}{{Physics Reports}}

\definecolor{red}{rgb}{0.8,0,0}
\definecolor{violet}{rgb}{0.4,0,0.4}
\definecolor{green}{rgb}{0,0.5,0.0}
\definecolor{navy}{rgb}{0.0,0.0,0.6}
\definecolor{orange}{rgb}{0.8,0.2,0.0}

\begin{document}

\title{Magnetar superconductivity versus magnetism: Neutrino cooling processes} 
\author{Monika~Sinha }
\thanks{Present address: Indian Institute of Technology Rajasthan, 
Old Residency Road, Ratanada, Jodhpur 342011, India
}
\affiliation{Institute for Theoretical Physics,
 J.~W.~Goethe-University, D-60438  Frankfurt am Main, Germany}

\author{Armen Sedrakian}
\affiliation{Institute for Theoretical Physics,
 J.~W.~Goethe-University, D-60438  Frankfurt am Main, Germany}

\date{\today}

\begin{abstract}
  We describe the microphysics, phenomenology, and astrophysical
  implication of a $B$-field induced unpairing effect that may occur
  in magnetars, if the local $B$-field in the core of a magnetar
  exceeds a critical value $H_{c2}$. Using the Ginzburg-Landau
  theory of superconductivity, we derive the $H_{c2}$ field for proton
  condensate taking into the correction ($\le 30\%$) which arises
  from its coupling to the background neutron condensate. The density
  dependence of pairing of proton condensate implies that $H_{c2}$ is
  maximal at the crust-core interface and decreases towards the center
  of the star. As a consequence, magnetar cores with homogenous
  constant fields will be partially superconducting for
  ``medium-field'' magnetars ($10^{15}\le B\le 5 \times 10^{16}$ G)
  whereas ``strong-field'' magnetars ($B>5\times 10^{16}$ G) will be
  void of superconductivity.  The neutrino emissivity of a magnetar's core
  changes in a twofold manner: (i)~the $B$-field assisted direct Urca
  process is enhanced by orders of magnitude, because of the unpairing
  effect in regions where $B\ge H_{c2}$; (ii)~the Cooper-pair breaking
  processes on protons vanish in these regions and the overall
  emissivity by the pair-breaking processes is reduced by a factor of
  only a few. 
\end{abstract}

\maketitle

\section{Introduction}\label{intro}

The protonic fluid in the cores of magnetized neutron stars is a
type-II superconductor, i.e., it supports the magnetic $B$ field by
forming quantized electromagnetic vortices with density $n_v =
B/\Phi_0$, where $\Phi_0 =\pi\hbar c/e$ is the quantum of flux, in the
field range $H_{c1}\le B\le
H_{c2}$~\cite{1969Natur.224..674B,*1981PhRvB..24.2533M,*1991ApJ...380..515M,*1995ApJ...447..305S}.  
The lower critical field $H_{c1}$ is the field strength at which the emergence
of the first vortex (flux tube) becomes energetically favorable. At
the upper critical field strength $H_{c2}$ the normal cores of the
vortices touch each other and superconductivity is destroyed.  The
$H_{c2}$ field is density-dependent and is given by
\be \label{eq:1}
H_{c2} =\frac{\Phi_0}{2\pi\xi_p^2} , 
\ee
where $\xi_p$ is the coherence length of the proton condensate, which
scales inversely with the pairing gap $\Delta$. The coherence length
appears in Eq.~\eqref{eq:1} because $H_{c2}$ is the field at which the
Larmor radius of protons becomes comparable to the size of a Cooper
pair $\sim \xi_p$. A field $B\sim H_{c2}$ disrupts the coherence among
the protons which form a Cooper pair and, therefore, destroys their
superconductivity.  For the proton superconductor in the cores of
neutron stars $10^{15}\le H_{c2}\le 10^{17}$~G, i.e., $H_{c2}$ is well
above the fields expected in the interiors of ordinary neutron stars
($B\sim 10^{12}$-$10^{13}$ G).

The inferred magnetic fields on the surfaces of magnetars are of the
order of $ 10^{15}$ G. Their interior $B$ fields are not known, but it
has been frequently conjectured that they are larger than the surface
field. The conjectured maximal $B$ field, which is consistent with the
virial theorem for self-gravitating magnetic equilibria, is of the
order $B_{\rm max }\simeq 10^{18}$~G. Because $B_{\rm max } > H_{c2}$
and because these fields may vary over the star's core, we may
anticipate an intimate interplay between the magnetism and
superconductivity in the interiors of magnetars depending on whether
the local field is above or below $H_{c2}$.  Some observational
arguments where put forward in recent years in favor of type-I
superconductivity~\cite{2003PhRvL..91j1101L,*2004PhRvL..92o1102B,*2004PhRvC..69e5803B,
*1997MNRAS.290..203S,*2005PhRvD..71h3003S,*2007PhRvC..76a5801C,2005PhRvC..72e5801A}. 
However, our choice of the equations of state of dense matter and
microscopic parameters of the proton superconductor predict type-II
superconductivity throughout  most of the core of a neutron star,
as we show below.

The purpose of this work is to show that large enough magnetic fields
in the interiors of magnetars unpair proton superconductor in a
strongly density-dependent manner. We then go on to study the
consequences of this magnetically induced unpairing effect on the
neutrino emissivity of magnetars. Neutrino emissivities are the key
ingredients for the simulations of thermal evolution of magnetars and
can be confronted with the measured x-ray fluxes from the surfaces of
magnetars. 

This paper is structured as follows. In Sec.~\ref{sec:physics_input}
we discuss the input physics, i.e., the underlying equation of state
(EoS) and composition of matter which sets the stage for the following
discussion.  In Sec.~\ref{sec:GL_theory}, starting from the
Ginzburg-Landau (GL) functional for proton superconductor coupled to
neutron superfluid, we derive an expression $H_{c2}$, which accounts
for the density-density coupling between the proton and neutron
condensates.  In Sec.~\ref{sec:emissivities} we compute the
neutrino emissivities of the Urca process and pair-breaking processes
in magnetars including the unpairing effect.  Our conclusions and an
outlook are given in Sec.~\ref{sec:outlook}.

\section{Microphysical input}
\label{sec:physics_input}

\begin{figure}[thb]
\begin{center}
\includegraphics[width=8cm]{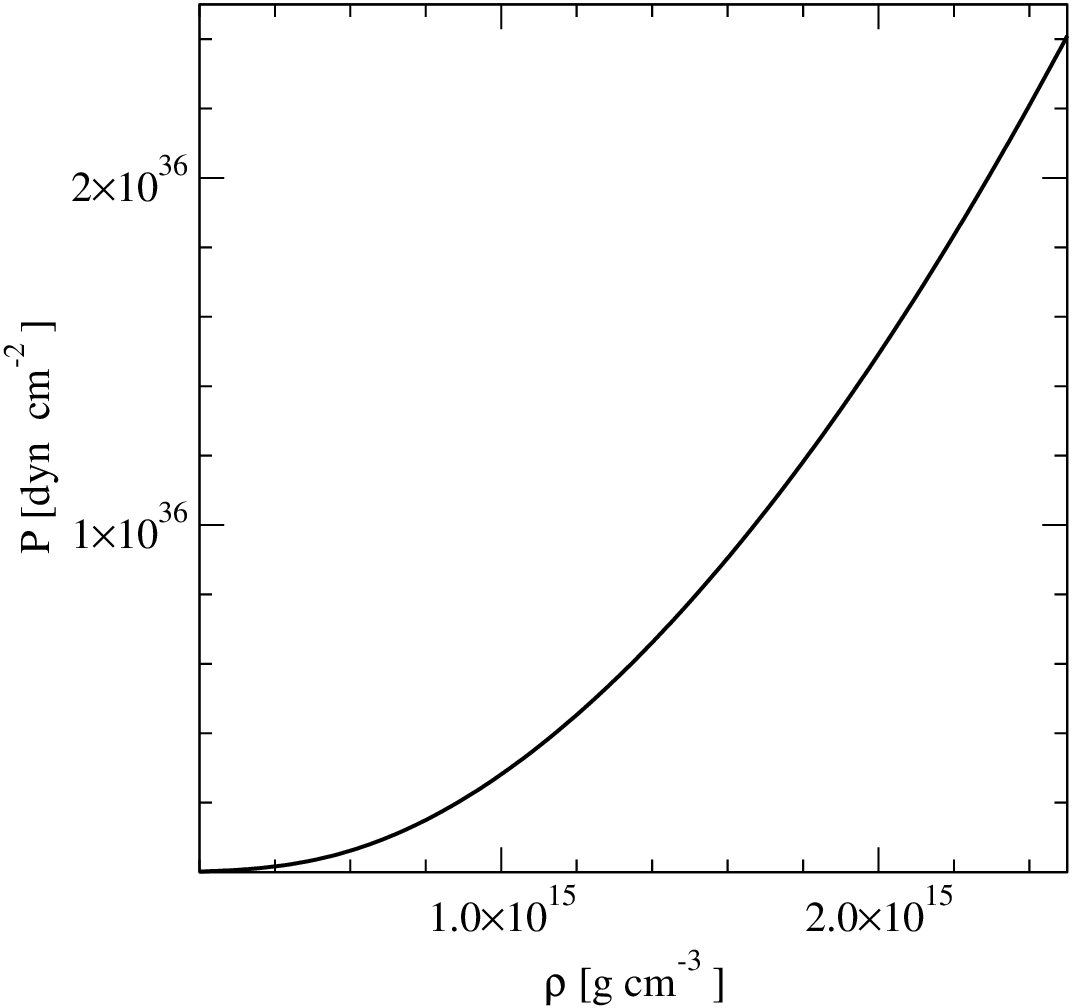}
\caption{Zero-temperature equation of state of dense matter composed
  of neutrons, protons, electrons, and muons in $\beta$ equilibrium
  derived from the relativistic DF with the parametrization of 
  Ref.~\cite{2005PhRvC..71b4312L}.}
\label{fig:EoS}
\end{center}
\end{figure}

Consider a magnetar with a nonstrange baryonic core consisting of
neutrons ($n$), protons ($p$), electrons ($e$), and muons ($\mu$) in
$\beta$ equilibrium. We choose to work with a relativistic density
functional (DF) with density-dependent couplings derived in
Ref.~\cite{2005PhRvC..71b4312L} to obtain the equation of state (EoS)
and composition of matter in the star's core and inner crust.  The
latter parametrization is in excellent agreement with the nuclear
phenomenology as it predicts saturation density $n_0=0.152$ fm$^{-3}$,
binding energy per nucleon $E/A=-16.14$ MeV, incompressibility
$K_0=250.90$ MeV, symmetry energy $J=32.30$ MeV, symmetry energy slope
$L=51.24$ MeV, and symmetry incompressibility $K_{sym} = -87.19$ MeV
all taken at saturation density~\cite{2011PhRvC..83d5810D}.  For
completeness we show the EoS of baryonic matter in Fig.~\ref{fig:EoS}
derived from this DF. Compact star models based on this DF were
constructed
elsewhere~\cite{2013PhRvC..87e5806C,*2014PhLB..734..383V,*2014JPhCS.496a2003C}
where it has been shown that the resulting maximum mass predicted by
this EoS is well above the current observational lower limit
2$M_{\odot}$ on the maximum mass of any compact star.  Strangeness in
form of hyperons or deconfined two- or three-flavor quark matter may
appear in the centers of magnetars, but we neglect this possibility in
the following.

\begin{figure}[tbh]
\begin{center}
\includegraphics[width=8cm]{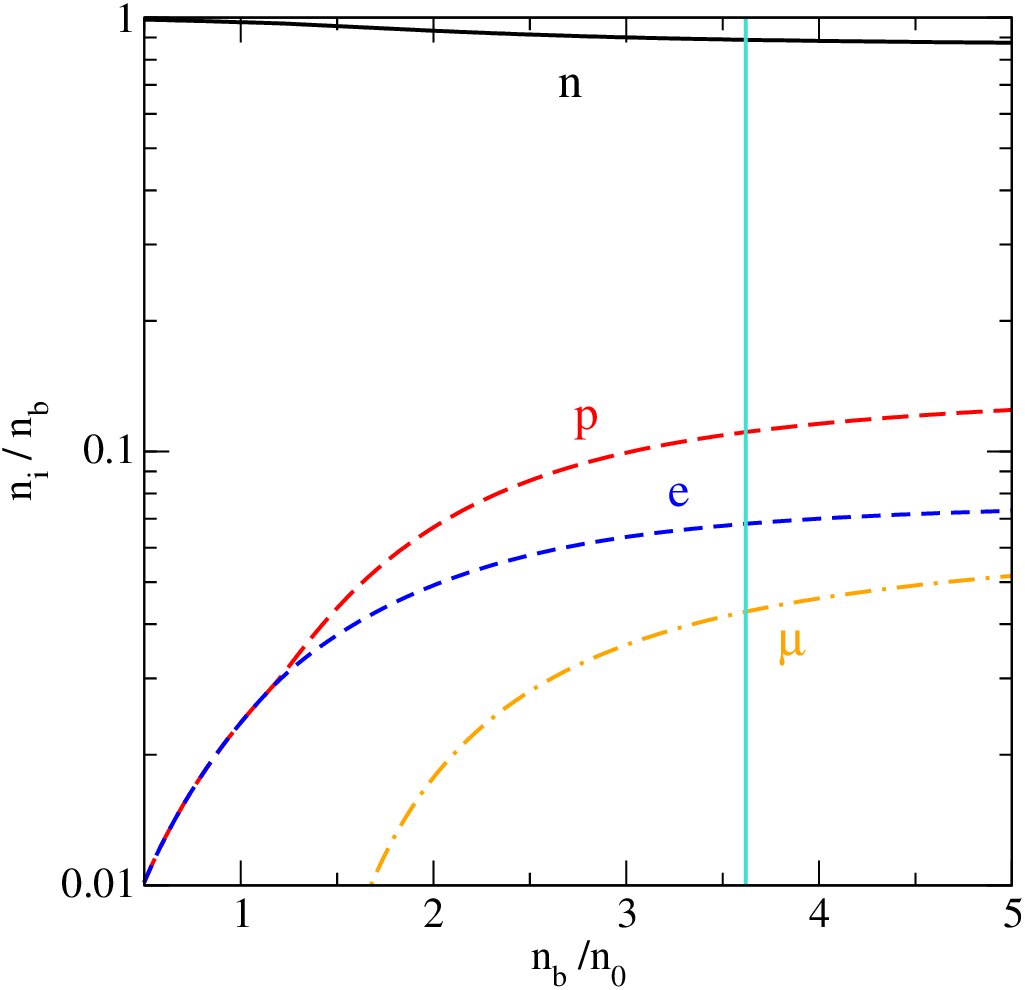}
\caption{ (Color online) Dependence of particle abundances $n_i/n_b$, $i\in n, p, e,
  \mu$ on the net baryon density $n_b$ in units of saturation density
  $n_0=0.152$ fm$^{-3}$. The vertical line shows the approximate Urca
  threshold $Y_{\rm Urca} = 0.11$ for proton fraction. }
\label{fig:abundances}
\end{center}
\end{figure}

The composition of dense matter corresponding to our EoS is shown 
in Fig.~\ref{fig:abundances}, where we show the abundances of species 
$n_i/n_b$, where $i\in n, p, e, \mu$ as a function of baryon density
$n_b$ normalized by the nuclear saturation density of the DF. 
\begin{figure}[tbh]
\begin{center}
\includegraphics[width=8cm]{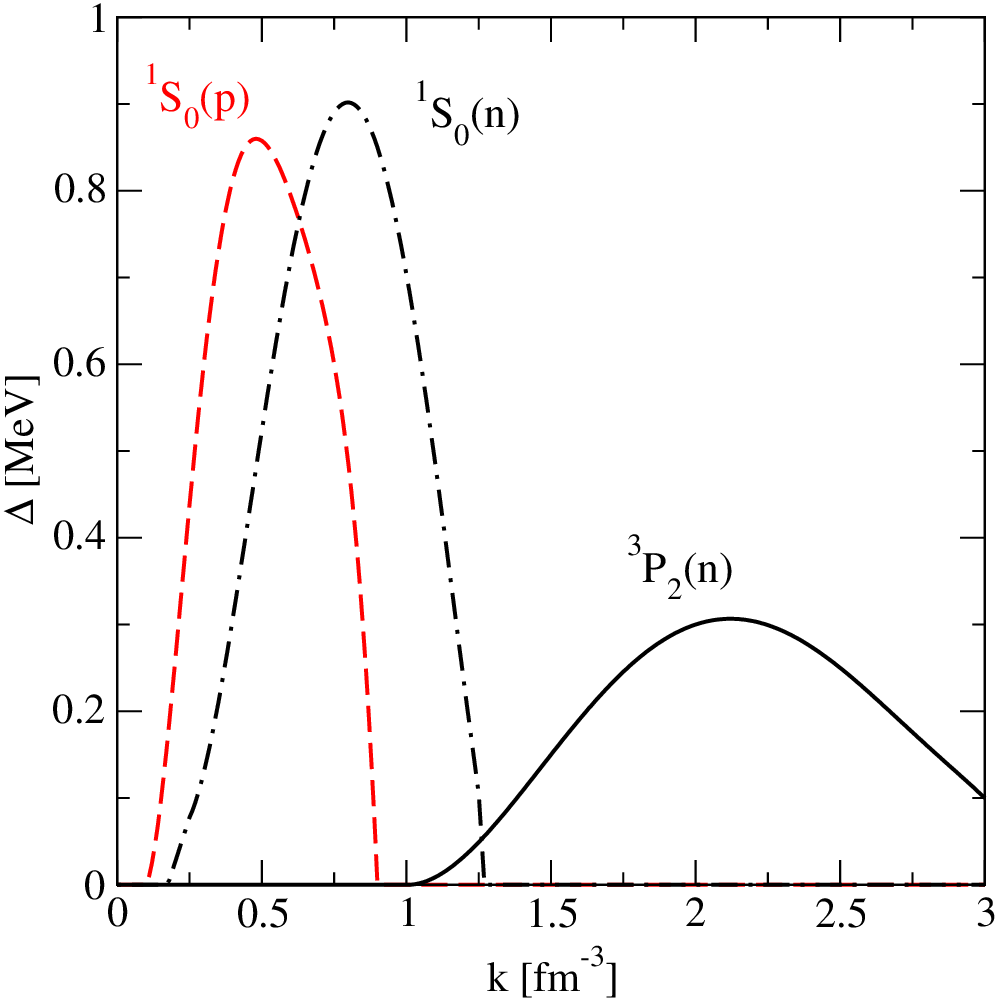}
\caption{(Color online) Dependence of $S$- and $P$-wave pairing gaps of neutrons
  (dash-dotted and solid lines) and of $S$-wave protons (dashed line) on their
  respective Fermi momenta.}
\label{fig:gaps}
\end{center}
\end{figure}
The abundances of protons and electrons are equal up to the point
where muons set in. The threshold value of Urca process in
non-magnetized matter $n_p/n_b\sim 0.11$ is reached at the density
$n \simeq 3n_0$. The composition of matter itself will
  be affected by a strong $B$ field, when electromagnetic interactions
  become of the order of the nuclear scale set by the Fermi energies
  of the constituents. However, below the field values $10^{18}$~G the
  abundances of baryons for non-zero $B$ are indistinguishable from
  those in the $B=0$ case (see Ref.~\cite{2013NuPhA.898...43S} and
  references therein).  

The pairing channels in neutron star matter correspond to the attractive
most dominant phase-shifts at given density or energy of nucleons (for
a review see, e.g., Ref.~\cite{2006pfsb.book..135S}).  Low density
neutron matter in the crust of compact stars pairs in the $^1S_0$
channel; above the saturation density the neutron fraction is large
enough (and energies are high enough) to render the $S$-wave
interaction repulsive. The dominant pairing at these densities is in
the $^3P_2$-$^3F_2$ channel.  Protons are in the continuum in the
fluid core of the star and are much less abundant than neutrons,
therefore their energies are low enough to favor the $^1S_0$ pairing.
The pairing gaps in these dominant channels adopted from
Refs.~\cite{1993NuPhA.555..128W,*1998PhRvC..58.1921B,*1992NuPhA.536..349B}
are shown in Fig.~\ref{fig:gaps}. The formulas which fit these gaps
are listed in the Appendix.  At asymptotically high densities pairing
of protons in the $^3P_2$-$^3F_2$ channel may occur. In the exceptional
models where matter is nearly isospin symmetrical at high densities,
spin-one, isospin-zero pairing in the $^3D_2$ channel may become the dominant
one~\cite{1996NuPhA.604..491A}.  Examples of such models are those
which feature kaon condensates.  Below we do not consider proton
$P$-wave pairing or neutron-proton $D$-wave pairing.

\section{Ginzburg-Landau theory of $H_{c2}$ in dense matter}
\label{sec:GL_theory}

Type-II superconductivity is characterized by the GL parameter,
$\kappa= \delta_L/\xi_p$, where $\delta_L$ is the London penetration
depth of the $B$ field in a superconductor, having the range
\be\label{eq:GL}
 \frac{1}{\sqrt{2}}  < \kappa <\infty.
\ee 
The critical value $\kappa_c=1/\sqrt{2} = 0.7071$ separates the domains
of type-I and type-II superconductivity.

The magnetic field is confined to electromagnetic vortices for field
values between the lower $H_{c1}$ and upper $H_{c2}$ critical
fields. If the $B$ field is lager than $H_{c2}$ it unpairs the Cooper
pairs and the material makes a transition to the normal state. The
phase transition from superconducting to the normal state in the
vicinity of $H_{c2}$ can be described in terms of the GL theory,
because the superconducting order parameter is small.  Note that in
the vicinity of $H_{c2}$ the superconducting order parameter is small
because of the large $B$ field and the GL expansion is valid not
only near the critical temperature $T_c$, but in the entire
temperature range $0\le T\le T_{c}$.

We start by writing down the GL functional for a superfluid neutron
and superconducting proton mixture
\bea\label{eq:GL}
{F}[\phi,\psi] &=& {F}_n[\phi] +\alpha\tau \vert \psi\vert^2 +\frac{b}{2}\vert
\psi\vert^4
+b'\vert \psi\vert^2 \vert \phi\vert^2\nonumber\\
&+&\frac1{4m_p}\Bigg\vert 
\left(-i\hbar \vecnabla - \frac{2e}{c} \vecA\right)\psi\Big\vert^2 + \frac{B^2}{8\pi},
\eea
where $\psi$ and $\phi$ are the proton and neutron condensate
wave-functions, $m_p$ is the proton mass, $\tau = (T-T_{cp})/T_{cp}$,
where $T_{cp}$ is the critical temperature of superconducting phase
transition of protons. Here $\alpha$ and $b$ are the coefficients of
the GL expansion for the proton condensate, $b'$ describes the
density-density coupling between the neutron and proton
condensates. This type of GL functional was analyzed initially to
study the current-current coupling between the neutron and proton
condensates~\cite{1980Ap.....16..417S} (the entrainment effect, see
~Ref. \cite{1976JETP...42..164A,*1991SvPhU..34..555S}). More recent
study of Ref.~\cite{2005PhRvC..72e5801A} discusses the density-density
coupling between the neutron and proton condensates and provides the
relevant microscopic expressions for the coefficient of the GL
functional.  The
effective vector potential can be decomposed as $\vecA = \vecA_{\rm
  em} + \vecA_{\rm ent }$, where the first term is the ordinary vector
potential of electromagnetism with $\vecB = \vecnabla \times
\vecA$. The second term is the ``entrainment'' vector potential
$\vecA_{\rm ent } = (\hbar c/e) [ (m_p^*-m_p)/m_p]\vecnabla\phi$,
where $m^*_p$ is the proton effective mass, see
Ref.~\cite{1980Ap.....16..417S}.  The entrainment effect describes
the current-current coupling between the neutron and proton
condensates.

The explicit form of the contribution of the neutron condensate to the
GL functional, ${F}_n[\phi]$, is not required in the following.

The minimization of the GL functional with respect to $\psi^*$, i. e., 
${\delta{F}[\phi,\psi]}/{\delta \psi^*} = 0 $
gives
\be
\label{eq:GL1}
\frac1{4m_p}\left(-i\hbar \vecnabla - \frac{2e}{c} \vecA\right)^2 \psi
+ \alpha \tau \psi + b|\psi|^2\psi + b^\prime|\phi|^2\psi = 0.
\ee 
The equilibrium value of the condensate is given by the solution of 
Eq.~\eqref{eq:GL1} 
\be
\label{eq:GL2}
 \psi(\alpha \tau  + b|\psi|^2 + b^\prime|\phi|^2) = 0,
\ee 
from which we obtain the two possible equilibrium solutions 
\bea
\psi &=& 0, \qquad\qquad  T> T_c,\\
\vert \psi\vert^2 &=& -\frac{1}{b}(\alpha\tau +
b'\vert\phi\vert^2),\quad T\le T_c.  
\eea 
The variation of the GL functional with respect to the electromagnetic
vector potential ${\delta{F}[\phi,\psi]}/\delta{\vecA} = 0 $ gives
\bea\label{eq:GL3}
 \frac{c}{4\pi}\vecnabla  \times \vecnabla \times \vecA = \vecj,
\eea
where 
\bea \label{eq:GL4}
\vecj  = -\frac{i\hbar
  e}{m} \left(\psi^*\vecnabla \psi- \psi\vecnabla
  \psi^*\right)-\frac{4e^2}{mc}\vert \psi\vert^2(\vecA_{\rm em}+2
\vecA_{\rm ent})\nonumber\\
\eea 
is the proton super-current.  It contains the conventional
electromagnetic current $\propto \vecnabla \psi$ as well as the
entrainment current $\propto \vecA_{\rm ent}\propto \vecnabla \phi$.

Equations \eqref{eq:GL1}, \eqref{eq:GL3},  and \eqref{eq:GL4} constitute
the GL equations in their most general form. To derive the value of
the upper critical field $H_{c2}$ it is sufficient to keep only
the linear in $\psi$ terms in the GL equations above.  To this order
Eq.~\eqref{eq:GL4} reduces to
\bea \label{eq:GL5} \vecnabla \times \vecnabla \times \vecA = 0 +
O(\vert\psi\vert^2).  \eea 
Furthermore, because $\vecnabla \times \vecA_{\rm ent} = 0$
identically (except at the singular points where the neutron vortices
are located), we can make the replacement $ \vecA \to \vecA_{\rm em} $
in Eq.~\eqref{eq:GL5}.  The small-scale (local) magnetic field is
homogenous, therefore the corresponding vector potential $\vecA_{\rm
  em}$ is linear in coordinates. We choose $\vecA_{\rm em}$ along one
of the directions of the Cartesian system of coordinates, say $z$
direction, without loss of generality.  Assume that the $B$ field is
in the $y$ direction. Then, $\psi = \psi(x)$ only. To linear order in
$\psi$ the solution of Eq.~\eqref{eq:GL2} is $ A_{\rm em} = Bx$.
Substituting this into the first GL equation \eqref{eq:GL1} one finds
\bea
\label{eq:oscillator}
- \psi''  + \frac{4\pi^2}{\Phi_0^2}B^2 x^2\psi
= - \frac{4m_p}{\hbar^2}\left(\alpha \tau +b^\prime|\phi|^2\right) \psi+ O(\vert\psi\vert^2) .\nonumber\\
\eea
The mathematical form of this equation is that of the harmonic
oscillator, therefore, its solutions is read off as  
\bea
\label{eq:harm_osc}
- \frac{4m_p}{\hbar^2}\left(\alpha \tau +b^\prime|\phi|^2\right) =
\left(n+\frac{1}{2}\right) \frac{4\pi  B}{\Phi_0}.
\eea
We are interested in the strongest field for which solutions with
$\psi \neq 0$ are still possible. This is the 
case $n=0$ in Eq.~\eqref{eq:harm_osc} which identifies  the critical field
$B=H_{c2}$. Consequently
\bea
\label{eq:Hc2_1}
H_{c2} &=& \frac{\Phi_0}{2\pi} \left[ -
 \frac{4m_p}{\hbar^2}\left(\alpha \tau +b^\prime|\phi|^2\right)
\right]\nonumber\\
&=&\frac{\Phi_0}{2\pi\xi_p^2} \left[1 + \frac{\vert b^\prime\vert
    |\phi|^2}{\alpha\vert\tau\vert} \right], 
\eea 
where we used the relation
$(m_p\vert\alpha\tau\vert)^{1/2} = \hbar /2\xi_p$ and the fact that
$b'<  0$, see below.  If $b'=0$ Eq.~\eqref{eq:Hc2_1} reduces to the
standard result~\cite{1980stph.book.....L}.  To evaluate the
correction to the critical field note that
$\vert \alpha\tau \vert= \vert\psi_0\vert^2 b$ and, therefore,
$( b'\vert\phi\vert^2/\alpha\tau) = ({n_n}/{n_p}) ({\vert
  b^\prime\vert }/{\vert b\vert}).  $

The coefficient $b'$ which takes into account beyond mean-field
coupling between the neutron and proton condensates was computed by
Alford {\it et al.} \cite{2005PhRvC..72e5801A} in $\beta$-equilibrated,
charge-neutral nuclear matter diagrammatically.  They also provide the
mean-field expression for $b$. Using their results we find that
\bea \label{eq:ratio}\frac{n_n}{n_p} \frac{\vert
  b^\prime\vert}{\vert b\vert} &=&\frac{27\pi^2}{4} G_{np}
\frac{n_n^2}{\mu_p^2\mu_n^2} \frac{\Delta_p^2}{m_pk_{F_p}}, 
\eea
where we used the value of $b'$ valid in the regime
$\Delta_p\ll \mu_p$ and $\Delta_n\ll \mu_n$ and $T\to 0$ and the value
of parameter
$G_{np} = 10^{-5}$ MeV$^{-2}$~\cite{2005PhRvC..72e5801A}.  Note that
close to the critical temperature $T_c$ alternative expressions
provided by Alford {\it et al.} \cite{2005PhRvC..72e5801A} should be used.
The correction in Eq.~\eqref{eq:Hc2_1} owing to the
  coupling between the neutron and proton condensates is $\le 30\%$;
  it is small because the coupling between the condensates arises only
  via fluctuations which vanish in the ground state.  The main
  uncertainty in Eq.~\eqref{eq:ratio} is the contact pairing
  interaction in the isosinglet channel $G_{np}$; its value quoted
  above should be viewed as an order of magnitude estimate. An
  additional uncertainty arises from the not well-known value of the
  gap in the proton spectrum $\Delta_p$ which may vary by a factor of
  few.

The analogy between Eq.~\eqref{eq:oscillator} and the one describing
harmonic oscillator in quantum mechanics can be exploited further to
write down the most-general ``harmonic oscillator'' type solution of
Eq. \eqref{eq:oscillator}, which describes a vortex in the $x$-$y$ plane
(with the field directed in the $z$ direction). The corresponding
wavefunction can be written as
\be
\psi(x,y) = \sum_{n=-\infty}^{\infty} C_n\exp[-\kappa B(x-k/\kappa
B)^2/2+iky], \ee 
where the coefficient $C_n$ and $k$ depend on the type of the proton
vortex lattice. Assuming triangular lattice one finds $k = \kappa
(\pi\sqrt{3} )^{1/2}$ and the set of condition $C_{n+4} = C_n$, $C_0 =
C_1 = C$, $C_2 = C_3 = -C$, where $C$ is given by the normalization of
the wave-function to the density of condensate. 
\begin{figure}[t]
\begin{center}
\includegraphics[width=8cm]{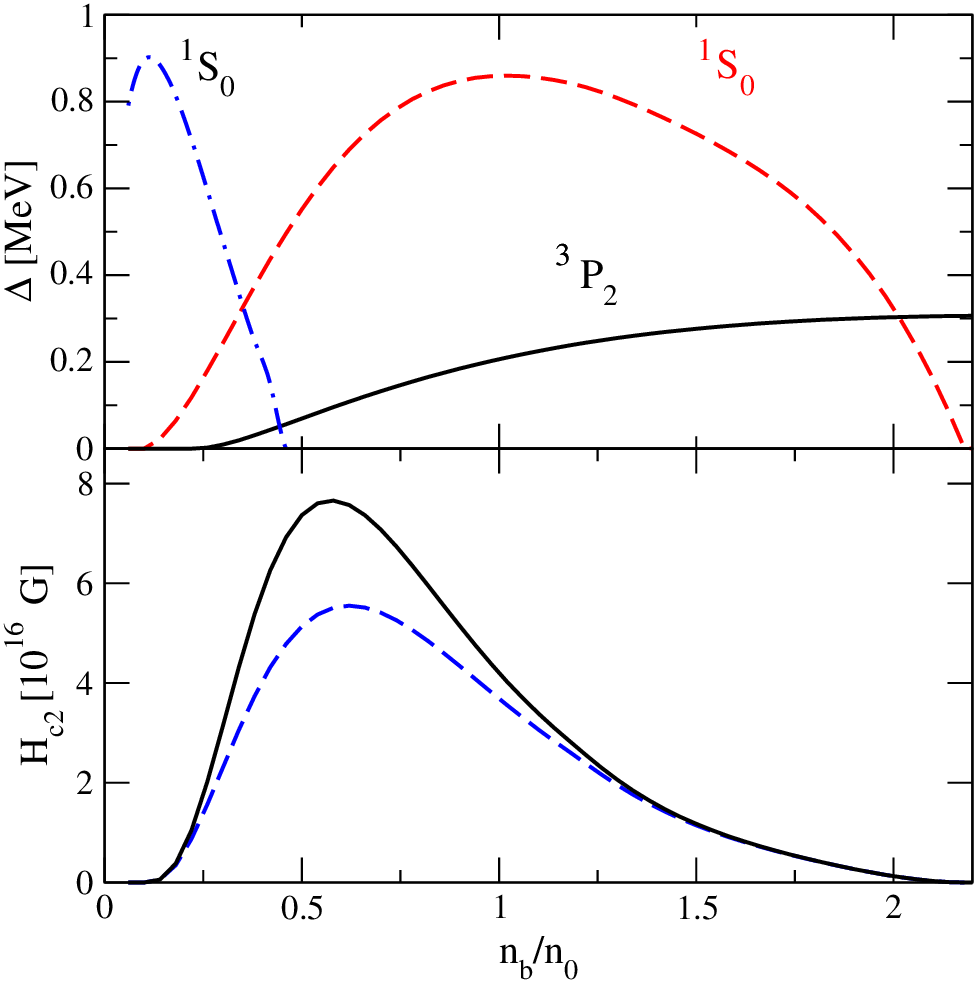}
\caption{ (Color online) Upper
 panel: Dependence of pairing gaps for neutrons
  ($^1S_0$ and $^3P_2$ channels) and for protons ($^1S_0$ channel) on
  baryonic density in units of nuclear saturation density. Lower
  panel: Dependence of the critical unpairing field $H_{c2}$ on
  baryonic density with account for the coupling between the neutron
  and proton condensates (full line) and without (dashed line). 
}
\label{fig:Hc2}
\end{center}
\end{figure}

Table I lists the key parameters of the proton superconductor for a
range of densities corresponding to the star's fluid core. The
coherence length has a minimum, which reflects the density dependence
of the gap ($\xi_p\propto \Delta^{-2})$. Because $H_{c2}$ scales
inversely with $\xi_p^{2}$, the critical field has a maximum, with
max~$ H_{c2} = 7.37\times 10^{16}$~G at $n_b = 0.7 n_0$ in our setup.
The London penetration depth scales as the root of inverse proton
density, therefore it decreases as the density increases. This has the
consequence that the GL parameter drops below the critical value
$\kappa_c$ and the proton superconductor becomes type-I. However, this
occurs only in the high-density end of the proton superconductivity domain
and should be relevant only for compact stars with central densities
exceeding this value.
\begin{table}
\begin{tabular}{ccccllll}
\hline
\\
$n_b/n_0$  &  $k_{Fp}$ & $\Delta_p$  & $m^*_p/m_p$	&  $\xi_p$ &
$\delta_L$ & $\kappa$ & $H_{c2}$\\
\\
\hline
 0.140 & 0.12 & 0.02 & 0.93 & 76.1 & 929.2 & 12.2 & 0.06 \\
 0.300 & 0.20 & 0.24 & 0.89 & 11.9 & 425.0 & 35.6 & 3.15 \\
 0.500 & 0.28 & 0.55 & 0.85 & 8.0 & 238.6 & 29.8 & 7.37 \\
 0.700 & 0.36 & 0.76 & 0.81 & 7.8 & 161.1 & 20.6 & 7.08 \\
 0.900 & 0.44 & 0.85 & 0.78 & 8.7 & 119.5 & 13.7 & 5.15 \\
 1.100 & 0.51 & 0.86 & 0.76 & 10.4 & 93.9 & 9.1 & 3.39 \\
 1.300 & 0.58 & 0.81 & 0.74 & 13.0 & 75.2 & 5.8 & 2.06 \\
 1.500 & 0.67 & 0.73 & 0.71 & 17.0 & 61.0 & 3.6 & 1.18 \\
 1.700 & 0.74 & 0.62 & 0.70 & 22.8 & 51.2 & 2.2 & 0.64 \\
 1.900 & 0.81 & 0.45 & 0.68 & 35.0 & 44.3 & 1.3 & 0.27 \\
 2.100 & 0.88 & 0.16 & 0.67 & 106.4 & 39.2 & 0.4 & 0.03 \\
\hline\\
\end{tabular}
\caption{Microscopic parameters of proton superconductor and the upper
critical field for unpairing $H_{c2}$ for a range of matter densities.}
\label{table1}
\end{table}

Figure~\ref{fig:Hc2} (upper panel) displays the dependence of pairing
gaps on baryon density for the composition of matter implied by our
chosen EoS. (Note that gaps displayed in Fig.~\ref{fig:gaps} as
functions of Fermi momenta of particles are EoS independent, whereas
those in Fig.~\ref{fig:Hc2} are specific to our EoS).  The dependence
of the $H_{c2}$ field on density is shown in Fig.~\ref{fig:Hc2} (lower
panel). It is seen that magnetars with interior fields with $B\le {\rm
  max}~H_{c2} \simeq 7.37\times 10^{16}$ G will be partially
superconducting, which means that regions where $B <H_{c2} $ will be
superconducting whereas the regions where $B >H_{c2} $ are not. Clearly,
magnetars with $B> {\rm max}~H_{c2} $ will be fully
non-superconducting. The maximum of $H_{c2} $ is attained close to the
crust-core interface (corresponding to $n_b = 0.5 n_0$). This implies
that for partially superconducting magnetars with $B< {\rm max}~H_{c2}
$ the unpairing by the magnetic field will remove proton
superconductivity in the inner core, whereas the outer core could be
still superconducting provided the $B$ field is approximately
homogeneous and constant in the fluid core of the star. This is a
reasonable assumption, because the density gradients are small in the
fluid core.

\section{Neutrino emissivity of magnetar cores}
\label{sec:emissivities}

This section studies the implications of the unpairing effect, discussed
in the previous section, on the neutrino emission processes from dense
matter in magnetars. We focus below on the neutrino emission processes
which are dominant below the critical temperature $T_{cp}$ of proton
superconductivity, specifically the $B$ field assisted Urca and the
pair-breaking processes. The implications of the unpairing effect for 
processes such as the modified Urca process and the modified
bremsstrahlung process are analogous to those for the direct Urca
process and the implementations in numerical codes should be
straightforward. 

\subsection{Direct Urca process}

The direct Urca process is kinematically allowed only above the
threshold $Y_{\rm Urca} = n_p/n_b > 11 \%$ in ordinary low-field
compact stars, because for low proton concentrations the energy and
momentum conservation cannot be fulfilled simultaneously
\cite{1991PhRvL..66.2701L,*1992RvMP...64.1133P,*1994PhR...242..297P}.
Strong $B$ fields change the phase-space of baryons.  As a
consequence, the direct Urca process is allowed even below the
threshold 
$Y_{\rm   Urca}$~\cite{1998PhRvD..58l1301B,*1998JHEP...09..020L,1999AA...342..192B}.
To characterize the kinematics of the Urca process in a $B$ field it
is convenient to introduce the parameter~\cite{1999AA...342..192B}
\be x = \frac{k_{Fn}^2 -
  (k_{Fe}+k_{Fp})^2}{k_{Fn}^2} N_{Fp}^{2/3}
\ee
where $N_{Fp} = k_{Fp}^2/2\vert e\vert B$ is the number of Landau
levels populated by protons. Thus, for $x>0$ the Urca process is
forbidden in the low-field limit, but can become operative in strong
magnetic fields. 

If, under such conditions, the Urca process operates
at a fraction of its strength, it can still be an important factor in
cooling the star's core, because other processes are by orders of
magnitude weaker. For $x<0$ the Urca process is allowed and the role
of the magnetic field is to induce ``de Haas--van Alfven'' type
oscillations in the emissivity of this process as a function of $B$ field.

The second effect of the strong magnetic field on the Urca process (not
discussed so far) is the effect of unpairing of the proton superconductor
by the field.  Proton and neutron pairings restrict the phase space
available for the process and, as a consequence, the rate of the
direct Urca process is suppressed. This suppression at asymptotically
low temperatures is given simply by an exponential quenching factor
$\exp(-\Delta/T)$ for each participating nucleon, where $\Delta$ is
the relevant pairing gap, $T$ is the temperature (more accurate
treatments are given, e.g., in Ref.
\cite{1999PhyU...42..737Y,*2007PrPNP..58..168S}). As outlined in
Sec.~\ref{sec:GL_theory}, large $B$ fields unpair the proton
superconductor, therefore the suppression of the Urca neutrino
emission due to the gap in proton quasiparticle spectrum will be
absent, i.e., only neutron pairing will contribute to the
suppression. Because the gap for neutrons in the $P$-wave channel is
smaller than the one in the $S$-wave channel for protons (see
Fig.~\ref{fig:Hc2}), the onset of suppression will strongly deviate
from the one expected in the case of superconducting protons.

We now illustrate these qualitative arguments by numerical
examples. In our setup the proton fraction remains below $Y_{\rm
  Urca}$ in the density range where proton $S$-wave superconductivity
exists, i.e., densities $n\le 3n_0$, therefore we explore first the
domain $x> 0$, where Urca process is forbidden in the zero-field
limit.

The Urca emissivity for $B\neq 0$ is written as~\cite{1999AA...342..192B}
\be \label{eq:urcaB}
\epsilon^{\rm Urca} = \frac{457\pi G_F^2}{10080} (1+3g_A^2)m_n^*m_p^*\mu_e
T^6 {\cal R} {\cal S}_n {\cal S}_p.
\ee
where $G_F$ is the Fermi coupling constant, 
$g_A$ is the axial-vector coupling, $m_{n/p}^*$ are the
effective masses of neutron and proton, $\mu_e$ is the chemical
potential of electrons, 
${\cal R}$ function encodes modifications due to the field and
${\cal S}_{n/p} = \exp (-\Delta_{n/p}/T)$ are the suppression factors
arising owing to the pairing of neutrons ($n$) and protons ($p$). The
quenching of proton superconductivity implies ${\cal S}_p = 1$ in
Eq.~\eqref{eq:urcaB}. To model the function ${\cal R}$ in the
forbidden region we use an approximate polynomial fit to the functions
shown in Fig. 1 of Ref.~\cite{1999AA...342..192B}, which
is given by
$\log_{10} {\cal R} = -0.35942 - 0.506418 x + 0.0130305 x^2 - 0.00140399 x^3.  $
In the allowed domain we use the fit formula~\cite{1999AA...342..192B}
\be
{\cal R} =1-  \frac{\cos \phi}{0.5816 + \vert  x\vert ^{1.192}}
\ee
where $\phi \equiv (1.211+0.4823\vert x\vert +0.8453\vert  x\vert^{2.533}) 
/(1+1.438\vert x\vert^{1.209})$ which is valid in the range $-20\le
x\le 0$ and $N_{Fp}\to \infty$.
\begin{figure}[t]
\begin{center}
\includegraphics[width=8cm]{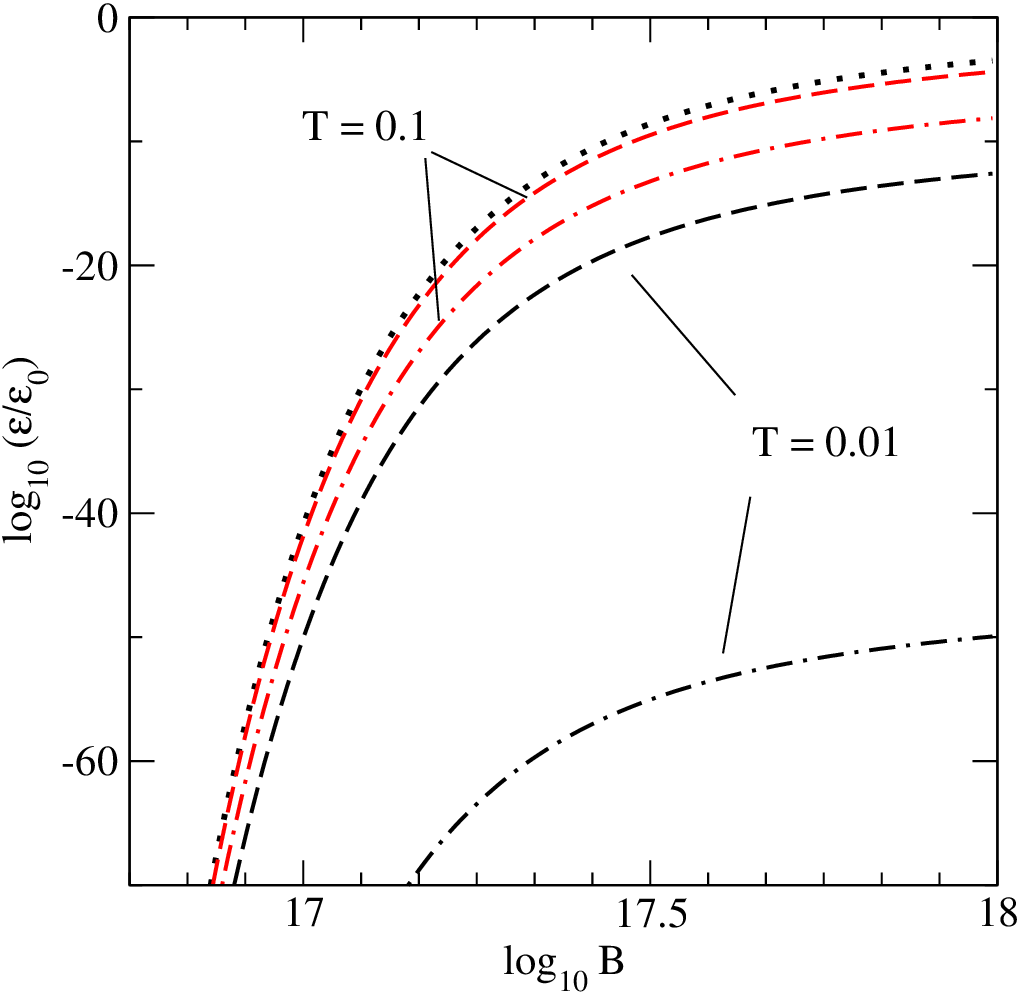}
\caption{ (Color online) The emissivity of the Urca process in the forbidden region
  ($x> 0)$ in units of the zero-field emissivity $\epsilon_0$ at fixed
  density $n =n_0$ for temperatures $T= 0.01$ and $0.1$ MeV. The Urca
  emissivity is shown for the cases of (a) normal matter (dotted
  line), (b) paired neutrons and normal protons (dashed lines) and (c)
  paired neutrons and protons (dashed-dotted lines). However the case
  (c) cannot be realized because of the unpairing effect for all the
  plotted values of $B > H_{c2} = 16.57 $. Note that for fixed density
  the scaling of the kinematical factor $x$ along the $B$-axis is
  given through its dependence on the number of Landau levels, i.e.,
  $x\propto N_{Fp}^{2/3}\propto B^{-2/3}$.  }
\label{fig:urca_forbid}
\end{center}
\end{figure}
Figure~\ref{fig:urca_forbid} displays the neutrino emissivity via the
Urca process in the forbidden region as a function of the $B$ field at
fixed density ($n = n_0$) and two values of temperature. The unpaired
case coincides with the results of
Refs.~\cite{1998PhRvD..58l1301B,*1998JHEP...09..020L,1999AA...342..192B}.
Magnetic field allows the Urca process to operate with emissivity
comparable with the emissivities of competing processes in the
asymptotically large field region $B\to B_{\rm max}$, as seen in
Fig.~\ref{fig:urca_forbid}. The pairing of neutrons and protons
requires an additional multiplicative factor ${\cal S}_n {\cal S}_p = 
\exp
[-(\Delta_n+\Delta_p)/T]$ in the neutrino emissivity. We show the
cases $\Delta_p = 0$ and $\Delta_p\neq 0$ assuming that the neutron
pairing gap $\Delta_n\neq 0$ and corresponds to its value at
$B=0$. Because for all $B$ field values $B> \textrm{max}\, H_{c2}$
(log$_{10} [\textrm{max} H_{c2}] = 16.87$) the unpairing effect requires $\Delta_p =
0$; thus the case $\Delta_p\neq 0$ is not realized physically, but
provides a measure of the error of neglecting the unpairing effect. It
is evident from Fig.~\ref{fig:urca_forbid} that this error is
substantial and is the consequence of the fact that $\Delta_p \gg
\Delta_n$ in our example. This condition holds except at the edge of
the density domain of interest, see Fig.~\ref{fig:Hc2}.  Thus, the
proton pairing, if allowed, would suppress the emissivity stronger
than the neutron pairing, but because of the unpairing effect the Urca
emissivity is suppressed by the neutron superfluidity only. As a
consequence the Urca emissivity would be enhanced from its value which
neglects the unpairing effect.
\begin{figure}[t]
\begin{center}
\includegraphics[width=8cm]{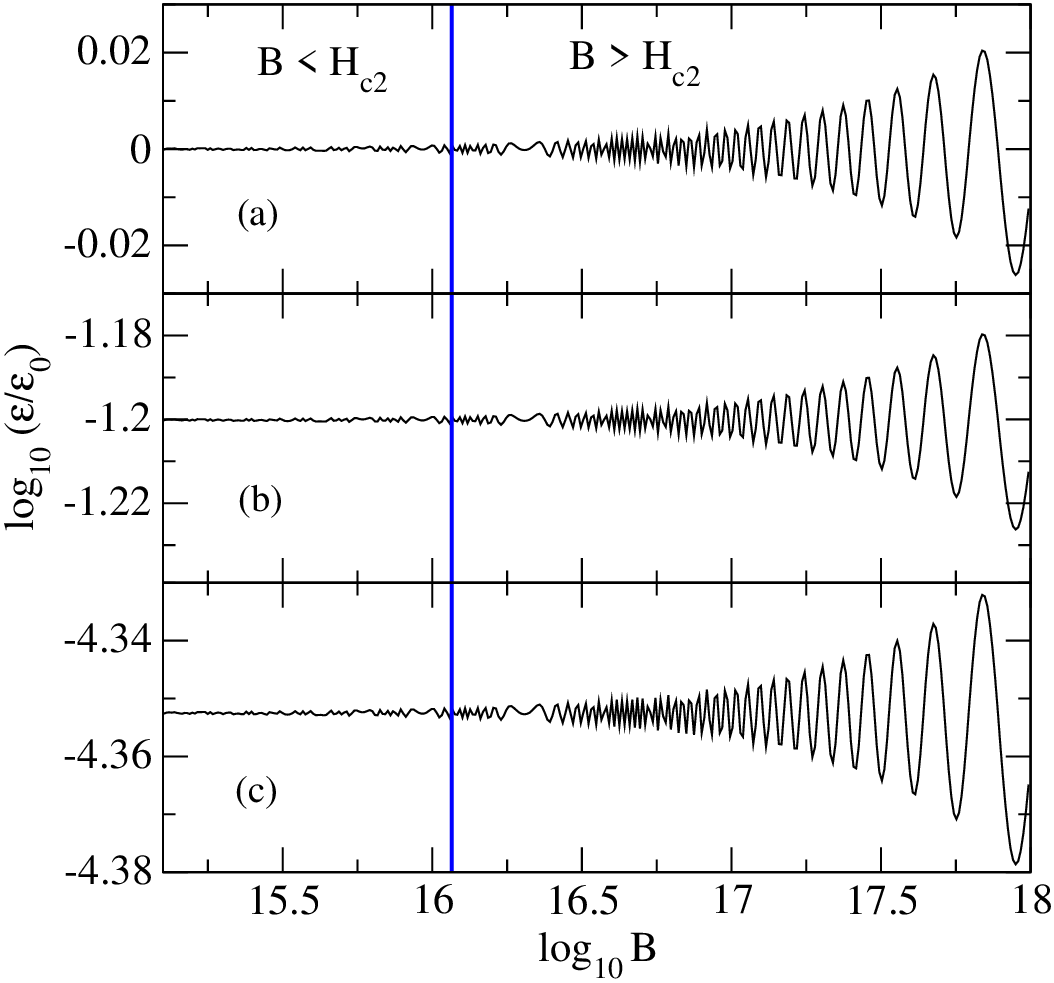}
\caption{ The emissivity of the Urca process in the allowed region
  ($x< 0)$ in units of the zero-field emissivity $\epsilon_0$ at fixed
  density $n =1.5\, n_0$ for temperature $T=0.1$ MeV. The magnitude of
  proton and neutron gaps are $\Delta_p = 0.73$ MeV and $\Delta_n =
  0.28$ MeV. The emissivity is shown in the cases of  (a) unpaired matter,
  (b) paired neutrons and for $B> H_{c2}$ (unpaired protons) and (c)
  paired neutrons and for $B< H_{c2}$ (paired protons).  }
\label{fig:urca_allowed}
\end{center}
\end{figure}
To explore the allowed region $x\le 0$ of kinematics for the Urca process
in strong magnetic field we have artificially increased the Fermi
momenta of protons and electrons by factor of two.  (In our models the
proton fraction exceeds the Urca threshold at density which is larger
than the maximal density at which proton $S$-wave superconductivity
exists).  We also choose to work at density $1.5\,
n_0$ because Urca process becomes operative in the high density domain.
  Figure~\ref{fig:urca_allowed} displays the neutrino
emissivity of the Urca process in the allowed region as a function of
the $B$ field for unpaired matter and for cases $B < H_{c2}$
(superconducting protons) and $B > H_{c2}$ (non-superconducting
protons). In the case of unpaired neutrons and protons (panel a), the
$B$ field induces de-Haas--van Alfven type oscillations in the
emissivity around its value in the zero $B$ field limit, as
expected~\cite{1998PhRvD..58l1301B,*1998JHEP...09..020L,1999AA...342..192B}.
For fields $B < H_{c2}$ the emissivity is suppressed by neutron and
proton pairing simultaneously; for $B > H_{c2}$ protons are unpaired
and the suppression is only due to paired neutrons. The transition
from one regime  to the other  is seen as a jump in the emissivity in
 (b) and (c) of Fig.~\ref{fig:urca_allowed} at
$B=H_{c2}$. The oscillations in (b) are around a value of
emissivity which is about an order of magnitude smaller than the
emissivity in the normal state, which reflects the suppression via
neutron pairing only. If the unpairing effect was neglected the emissivity
would have remained about 4--5 orders of magnitude below $B=0$ case.

 Strong magnetic fields will influence the neutron
  superfluidity in the curst ($S$-wave) and in the core ($P$-wave)
  differently. The $S$-wave condensate forms spin-zero Cooper pairs
  and the Pauli paramagnetic alignment of neutron spins along the
  $B$ field will act to quench their pairing. Generally, this
  quenching is effective for fields larger than those discussed here
  ($B> 10^{17}$ G), but the value of the critical field depends on the
  gap in the zero field limit, which has an uncertainty of an order of
  magnitude. The $P$-wave condensate forms spin-one Cooper pairs and
  the magnetic field will align the spins of Cooper pairs without
  affecting their internal structure. Initial studies of $P$-wave
  pairing in strong fields show that there is no suppression of the
  pairing induced by the field~\cite{2012JPhCS.400c2101T}. Therefore,
  as far as the Urca process is concerned, we do not expect additional
  suppression of pairing due to the $B$ field in the $P$-wave paired
  core.

  Thus we conclude that the unpairing effect which
  destroys the proton condensate can strongly influence the neutrino
  emissivity via the Urca process in the cores of magnetars. These
  modifications may have important consequences on the modeling of
  thermal transients and cooling in magnetars. 

There exists an additional channel of neutrino losses, which arises
once the interaction energy of the $B$ field with the spin of a
nucleon becomes of the order of temperature - the direct
bremsstrahlung process $N\to N+\nu+\bar\nu$, where $N$ refers to a
nucleon. This process is strictly forbidden in the non-magnetic case,
but becomes operative in a strong enough $B$ field,  because 
of the paramagnetic splitting of the energies of nucleons in a strong $B$
field. Spin-flip neutrino emission is effective within the window of
splitting energies of the order of the temperature of 
ambient matter~\cite{2000AA...360..549V}. By the same argument as in the case
of the Urca process above, the bremsstrahlung process $p\to p+\nu+\bar\nu$
will remain intact in magnetars, in contrast to the case of ordinary
neutron stars, where it would be suppressed by the proton pairing at
low enough temperatures.

\subsection{Pair-breaking processes}
The formation of nucleonic BCS condensates leads to the pair-breaking
 neutrino emission from each nucleonic condensate 
\cite{1976ApJ...205..541F,2006PhLB..638..114L,2006PhLB..638..114L,
  2007PhRvC..76e5805S,*2012PhRvC..86b5803S,2008PhRvC..77f5808K,
*2010PhRvC..81f5801K,2009PhRvC..79a5802S}.
In this subsection we study how the unpairing effect changes the neutrino
emissivity if the $B$ field exceeds the critical value $H_{c2}$ locally.
\begin{figure}[tbh]
\begin{center}
\includegraphics[width=8cm]{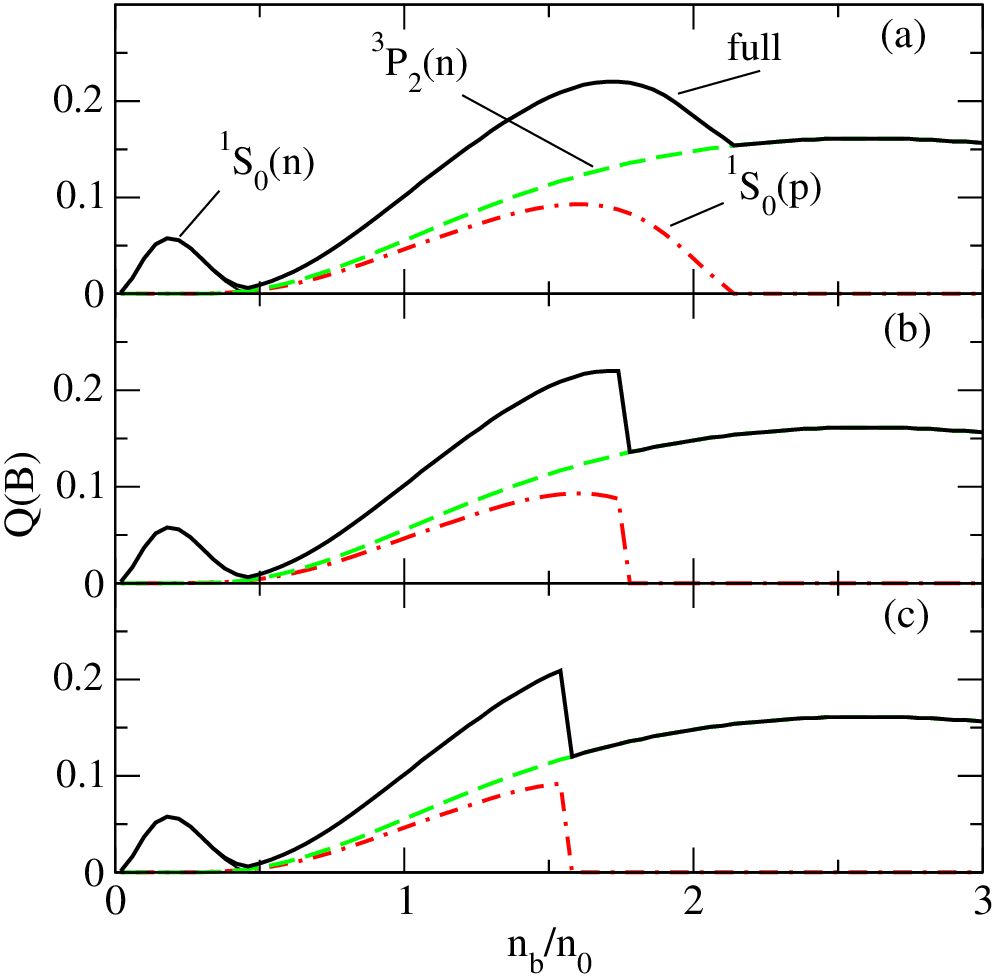}
\caption{ (Color online) Neutrino emissivity via pair-breaking
  processes as a function of baryon density in units of $n_0$ for
  $B_{16} = 0$ (a), $B_{16} = 0.5$ (b), and $B_{16} = 1$ (c).  The
  full pair-breaking emissivity is shown by solid lines and consists
  of $^1S_0$ neutron pair emission for $n_b/n_0 > 0.5$ and of the sum
  of $^1S_0$ proton and $^3P_2$ neutron pair emission for larger
  densities. The separate contributions of $^1S_0$ proton and $^3P_2$
  neutron pairs are shown by dash-dotted and dashed lines,
  respectively.  }
\label{fig:PB_emiss}
\end{center}
\end{figure}
To quantify the pair-breaking neutrino emission, consider their neutrino
emissivity, which is given by ($\hbar = c= 1$, our notations 
follow Ref.~\cite{2009ApJ...707.1131P})
\bea
\label{eq:PB_n}
\epsilon_{n} &=& \frac{4G_F^2m_{n}^* k_{Fn}}{15\pi^5} T^7
a_{n}^{S/P}\left(\frac{\Delta_n^{S/P}}{T}\right)^2{\cal I},\\
\label{eq:PB_p}
\epsilon_{p} &=& \frac{4G_F^2m_{p}^* k_{Fp}}{15\pi^5} T^7
a_{p}^{S}\left(\frac{\Delta_p}{T}\right)^2{\cal I},
\eea
where the subscripts $n$ and $p$
refer to neutrons and protons and the superscripts $S$ and $P$ refer to
$^1S_0$ and $^3P_2$ pairing of neutrons.  $\Delta_n^{P}$ in
Eq. (\ref{eq:PB_n}) stands for the angle averaged value of the
spin-triplet neutron gap, in which case it can
be factored out of the integral ${\cal I}$. (The explicit form of the
integral  ${\cal I}$  is not needed here and can be found, for
example, in Ref.~\cite{2009ApJ...707.1131P}).
 The $a$-coefficients are
defined as
\bea
a_n (^1S_0)&=& \frac{4}{81} c_{nV}^2v_{Fn}^4
+ \frac{11}{42}c_{nA}^2 v_{Fn}^2 \chi_n, \\
a_p (^1S_0)&=&  \frac{4}{81} c_{pV}^2v_{Fp}^4
+ \frac{11}{42}c_{pA}^2 v_{Fp}^2 \chi_p, \\
a_n (^3P_2)&=&\frac{c_{nA}^2}{2} ,
\eea 
where $\chi_{n/p}= 1+ (42/11) (m_{n/p}^*/m_{n/p})^2$, $c_{nV} = 1$, $C_{nA} =
g_A$, $C_{pV} = 4 \sin^2 θ_W - 1$, and $C_{pA} = −g_A$, with
$g_A\simeq 1.26$ and $\sin^2 \theta_W = 0.23$.  

Figure~\ref{fig:PB_emiss} displays the functions
\bea
Q_{n/p}(B) [\textrm{fm}^{-1}] = \frac{m_{n/p}^* k_{Fn/Fp}}{m_{n/p}}
a_{n/p}^{ S/P}\left(\frac{\Delta_n^{S/P}}{T}\right)^2,
\eea
which are more convenient for our analysis than the emissivities as
all common factors appearing in the emissivities \eqref{eq:PB_n} and
\eqref{eq:PB_p} are discarded (including the temperature, which is
assumed to be constant throughout the core and the inner crust of the
star).  In the crust of the star (i.e. for densities $n\le 0.5n_0$)
the pair-breaking emission is due to the $^1S_0$ paired neutron Cooper
pairs. This process is unaffected by the unpairing effect and is shown
for comparison. At larger densities, in the core of the star, neutron
and proton Cooper pair-breaking processes contribute about equally to
the neutrino energy loss in the zero-field limit [Fig. 7 (a)]. The
influence of the unpairing effect is seen in (b) and (c) where we
assume constant value of the field $B_{16} = 5\times 10^{15}$ 
and $B_{16} = 10^{16}$ G.  The constant $B$ field
removes the proton pair-breaking processes in the regions where
$B>H_{c2}$ locally, because the condensate vanishes in that
region. As a consequence the total emission rate is reduced to its
value corresponding to the emission by the $P$-wave condensate.

The unpairing effect will influence, apart from the emissivities of
the magnetars, also their thermal inertia, because the absence of
proton superconductivity will enhance the heat capacity of the
star. As a consequence the timescale needed for the star's
temperature to reach a given value will be larger than in the case of
absence of unpairing. In superconducting stars the main source of heat
capacity are electrons; in magnetars non-superconducting protons will
approximately double the heat capacity of the core of the star.  Thus,
we  anticipate that the proton and electron specific heats decrease
linearly with temperature as in normal Fermi liquids, whereas the heat
capacity of superfluid neutrons will be reduced by their superfluidity
(exponentially in the case of $S$-wave pairing and as power-law in the
case of $P$-wave pairing). The unpairing induced reduction of the
neutrino emissivity and the increase of the specific heat of matter
will both act to increase the cooling time-scale of the star.

\subsection{Relating the surface and crust-core boundary $B$ fields}
\label{sec:BsBrelation}
Because only the surface $B_s\simeq 10^{15}$ G fields are observed in
magnetars it appears to us useful to address the problem of relating
these observed surface fields to those  in magnetar interiors as
predicted by theoretical models.  Equilibria of magnetized neutron
stars with superconducting cores have been constructed in
Refs.~\cite{2008MNRAS.383.1551A,*2013MNRAS.431.2986H,2013PhRvL.110g1101L,*2014MNRAS.437..424L}.
Both poloidal and toroidal fields, as well as their combinations have
been considered. These studies suggest a linear relation of the form
\be
\label{HW2013} B_s \simeq \alpha_B H_b,
\ee
where $H_b$ is the field intensity at the outer boundary of the core
and $B_s$ is the surface field.  For purely poloidal field
Ref.~\cite{2013MNRAS.431.2986H} finds $\alpha_B = \epsilon_b/3$ where
$\epsilon_bR$ is the thickness of the crust, $R$ being the radius of
the star. This relation was derived for low fields $B\sim H_{c1}$,
where the role of the lattice of flux tubes can be neglected and these
can be treated as isolated entities. Its validity for larger fields
$B\sim H_{c2}$, more relevant to our discussion of the unpairing
effect, is not known. Nevertheless, we extracted the values of
$\epsilon_b$ using our EoS shown in Sec.~\ref{sec:physics_input}
assuming that the crust-core boundary is at $n\simeq 0.5n_0$. We find
that there is roughly two orders of magnitude drop in the field value
between the crust-core boundary and the surface of the
star. Specifically, for the 1.4 $M_\odot$ star model $\alpha_B =
0.058$ and for the 2.67 $M_\odot$ star model $\alpha_B =
0.021$~\cite{2014arXiv1403.2829S}. The study of
Ref.~\cite{2014MNRAS.437..424L}, which uses a different method,
suggests that the drop of the field from the magnetic pole to the base
of the crust is smaller and in the limit of large fields is of order
of unity. Clearly, further work is needed to establish the relation
\eqref{HW2013} in the strong-field regime $B\sim H_{c2}$. 
While the relation \eqref{HW2013} is highly important for relating
the physics of the unpairing effect to the observations of magnetars, our
discussion and results are independent of the value of the coefficient
$\alpha_B$ appearing in that relation.

\section{Summary and outlook}
\label{sec:outlook}

We have calculated the critical field $H_{c2}$ for unpairing for the proton
condensate, including its coupling to the density of the background
neutron condensate, using Ginzburg-Landau theory in the vicinity 
of superconducting-normal phase transition. We find that this coupling
enhances the value of the critical field by $\simeq 30\%$ (Fig.~\ref{fig:Hc2}).

The composition of dense matter and the dependence of proton pairing
gap on the Fermi momentum implies that the coherence length has a
minimum as a function of density which translate into a maximum in the
critical field (see Table I). The maximum is at the crust-core boundary
and the critical field decreases towards the center of the star.
Assuming the homogeneous constant $B$ field across the core and the inner
crust of the star, implies that magnetars with interior fields $B <
\textrm{max} ~H_{c2}$ are partially non-superconducting, whereas
magnetars with $B > \textrm{max} ~H_{c2}$ are void of proton
superconductivity.

The unpairing effect implies that the emissivity of the direct Urca
process is only Boltzmann-suppressed due to neutron gap and therefore
is more efficient than its counterpart in low-field matter, where
there is an additional suppression due to proton pairing (see
Figs.~\ref{fig:urca_forbid} and \ref{fig:urca_allowed}, which
illustrate the argument in the allowed and forbidden kinematical
domains, respectively). Unpairing further implies that the Cooper pair-breaking
processes in protonic matter are absent; this reduces the local net
pair-breaking emissivity of matter by a factor of a few (see
Fig.~\ref{fig:PB_emiss}).  In addition unpairing increases the
specific heat of magnetar cores and, therefore, the thermal inertia of
the core by a factor of two. Combined, the decrease in pair-breaking 
neutrino emissivity and the increase of the specific heat 
will increase the cooling time-scale of the star. This would be
counterbalanced by enhancement in the direct Urca cooling in the
strong field limit. Detailed cooling simulations can reveal the
relative importance of these different factors in the cooling of
magnetars, which we have discussed separately.

It is not possible to state firmly whether the unpairing effect is
operative in magnetars with observed surface $B$ fields $10^{15}$ G or
not, because the topology and the strength of the interior fields are
not known accurately. If there is an increase of the field from the
surface towards the center of the star (say by a factor of 10 to 15),
as suggested by a number of studies and broadly conjectured in the
literature, then the unpairing effect implies that the observed
magnetars are either partially or completely non-superconducting. 

A separate issue, to be studied further, is the influence of the
strong magnetic fields on the pairing in neutron matter. The $S$-wave
neutron pairing will be suppressed by strong magnetic field due to the
Pauli paramagnetic alignment of neutron spins along the $B$ field. The
Chandrasekhar-Clogston limiting field for the quenching of $S$-wave
neutron superfluidity is close to the limiting fields compatible with
gravitational equilibrium.  On the other hand, the $P$-wave paired
neutron fluid does not experience suppression in the $B$
field~\cite{2012JPhCS.400c2101T}. 

\section*{Acknowledgments}

The work of M.S. was supported by the Alexander von
Humboldt-Stiftung. A.S. was partially supported by a collaborative
research grant of the Volkswagen Foundation (Hannover, Germany) and
the Deutsche Forschungsgemeinschaft (Grant No. SE 1836/3-1).  We are
grateful to Mark Alford, John Clark, Nicolas Chamel, Chris Pethick,
and Ira Wasserman for helpful feedback on early versions of this work.

\appendix*
\section{Fitting formulas}
 The pairing gap for neutrons in the core and the crust and for protons
 in the core of the star were fitted by suitable functions (a sum of a
 polynomial and an exponential function) which depend on the
 Fermi-momenta of respective particles at zero temperature. These are
 given by 
\bea
\Delta_n( ^1S_0) &=&  2.76991 - 2.17347/\exp(k_{Fn}^2) - 5.91497k_{Fn}\nonumber\\
&+&17.653k_{Fn}^2 - 19.1544k_{Fn}^3  + 6.14977k_{Fn}^4, \nonumber\\
\eea
for neutron superfluid in the crusts
\bea
\Delta_n( ^3P_2) &=& 5.97989
-2.45018/\exp(k_{Fn}^2)-9.76221 k_{Fn}\nonumber\\
&+&6.24521k_{Fn}^2-1.73691 k_{Fn}^3+0.173889 k_{Fn}^4,\nonumber\\
\eea
for the neutron superfluid in the core and using the CD Bonn
interaction with Bruckner-Hartree-Fock spectrum, and
\bea
\Delta_p ( ^1S_0) 
 &=& -302.669 + 302.982/\exp(k_{Fp}^2) - 8.3717k_{Fp} \nonumber\\
&+& 369.944k_{Fp}^2 - 160.227k_{Fp}^3 - 11.3246k_{Fp}^4, \nonumber\\
\eea 
for the superconducting protons in the core.
 The effective masses of neutron and protons were assumed to
 be qual and given by the following fit formula 
\bea \frac{m^*}{m} &=&
 1.00661 - 0.649838k_F + 0.34416k_F^2\nonumber\\
&-&0.0441441k_F^3,
\eea
where $k_F$ stands for neutron or proton Fermi momentum. More accurate
treatment would require different effective masses for neutrons and
protons, but the corrections to the emissivities are expected to be
small, in the range of a few percent. 

\end{document}